\documentclass[twocolumn]{emulateapj}

\usepackage{amsmath, amsthm, amssymb}

\shortauthors{Lim, Carey \& Tan}

\begin{document}

\title{Spectroscopic infrared extinction mapping as a probe of grain growth in IRDCs}

\author{Wanggi Lim}
\affil{Depatment of Astronomy, University of Florida, Gainesville, FL 32611, USA}
\affil{Infrared Processing Analysis Center, California Institute of Technology, Pasadena, CA 91125, USA}

\author{Sean J. Carey}
\affil{Infrared Processing Analysis Center, California Institute of Technology, Pasadena, CA 91125, USA}

\author{Jonathan C. Tan}
\affil{Departments of Astronomy \& Physics, University of Florida, Gainesville, FL 32611, USA}

\begin{abstract}
We present spectroscopic tests of MIR to FIR extinction laws in IRDC
G028.36+00.07, a potential site of massive star and star cluster
formation. Lim \& Tan (2014) developed methods of FIR extinction
mapping of this source using {\it Spitzer}-MIPS $24\:\micron$ and {\it
  Herschel}-PACS $70\:\micron$ images, and by comparing to MIR {\it
  Spitzer}-IRAC $3$--$8\:\micron$ extinction maps, found tentative
evidence for grain growth in the highest mass surface density
regions. Here we present results of spectroscopic infrared extinction
(SIREX) mapping using {\it Spitzer}-IRS (14 to $38\:\micron$) data of
the same IRDC.  These methods allow us to first measure the SED of the
diffuse Galactic ISM that is in the foreground of the IRDC. We then
carry out our primary investigation of measuring the MIR to FIR
opacity law and searching for potential variations as a function of
mass surface density within the IRDC. We find relatively flat,
featureless MIR-FIR opacity laws that lack the $\sim12\:\micron$ and
$\sim35\:\micron$ features associated with the thick water ice mantle
models of Ossenkopf \& Henning (1994). Their thin ice mantle models
and the coagulating aggregate dust models of Ormel et al. (2011) are a
generally better match to the observed opacity laws. We also find
evidence for generally flatter MIR to FIR extinction laws as mass
surface density increases, strengthening the evidence for grain and
ice mantle growth in higher density regions.
\end{abstract}


\section{Introduction}

Massive star and star cluster formation are key processes for
understanding galaxy evolution, since massive stars are vital sources
of feedback for regulating the interstellar medium (ISM) and further
star formation activity, while star clusters are the basic building
blocks of galactic stellar populations. Despite this importance, we
understand massive star and star cluster formation only very poorly
because the regions forming such objects are relatively rare and thus
typically far away (at least a few kpcs for most Galactic sources) and
also deeply embedded inside dense molecular clouds. 

A better understanding of massive star and star cluster formation
should result if we can find and characterize the initial conditions
of these processes. Two key methods to achieve this involve studying the
amount of IR extinction and sub-mm/mm emission of dust grains in cold,
dense regions of molecular clouds in order to measure mass surface
densities and thus masses of cloud structures (see, e.g, Tan et
al. 2014, for a review, hereafter T14). To do this accurately, one
needs to know the opacities and emissivities of the dust grains,
including potential systematic evolution in these properties. This is
the goal of our study. A better understanding of dust properties will
also be important for modeling heating and cooling in the clouds,
astrochemistry, and the ionization fraction, which affects dynamics by
enabling coupling of the mostly neutral gas to magnetic fields.

The NIR to FIR properties of dust extinction (from 1 to 25~$\rm \mu
m$) have been measured in the Lupus molecular cloud using background
stars by Boogert et al. (2013). They found evidence for water ice
(presumably forming on dust grains) on sightlines with $A_V\gtrsim
2$~mag and that with increasing $A_K$ the $>5\:{\rm \mu m}$ continuum
extinction increases relative to $A_K$. The most extincted line of
sight that they probed had $A_K=2.46$~mag ($A_V\sim20$~mag). These are
relatively low mass surface density conditions compared to the clouds
that are thought to form massive stars and star clusters (T14). Lutz
et al. (1996) studied the MIR extinction law (from 2.5 to 9~$\rm \mu
m$) towards the Galactic Center using hydrogen recombination
lines. They found a relatively flat extinction law, e.g., compared to
bare grain models of Draine (1989). This sightline has $A_V\sim
30$~mag, but is essentially probing diffuse ISM conditions that are
not likely to be close to initial conditions for star
formation. McClure (2009) studied 5 to 20~$\rm \mu m$ dust extinction
properties on sightlines through molecular clouds with $0.3 \leq A_K <
7$ towards background stars, finding systematic evolution with
increasing $A_K$ that was attributed to ice-mantle mediated grain
coagulation (see also Flaherty et al. 2007; Chapman et al. 2009; c.f.,
Ascenso et al. 2013, who did not find evidence for extinction law
variations up to $A_V\sim 50$~mag).
 
Infrared dark clouds (IRDCs) (e.g., P\'erault et al. 1996; Carey et
al. 1998; Ormel et al. 2005; Rathborne et al. 2005; T14) cast shadows
at MIR wavelengths, $\sim10\:\micron$. They are occasionally opaque at
far-infrared wavelengths, up to $\sim100\:\micron$, against the
Galactic diffuse background emission (e.g., Lim \& Tan 2014, hereafter
LT14) due to their very high mass surface densities $\sim 1\:{\rm
  g\:cm^{-2}}$ ($A_V\sim200$~mag) and cold $\lesssim 15$~K
temperatures. Being massive clouds with such physical conditions,
IRDCs are likely to be representative of the initial conditions for
massive star and star cluster formation (T14).

Butler and Tan (2009, 2012; hereafter BT09, BT12) studied mass surface
densities, $\Sigma$, of 42 starless cores inside 10~IRDCs via MIR
extinction (MIREX) mapping by utilizing {\it Spitzer}-IRAC
$8\:\micron$ data of the GLIMPSE survey (Churchwell et al. 2010).
Butler, Tan and Kainulainen (2014, hereafter BTK14) used deeper {\it
  Spitzer}-IRAC $8\:\micron$ images toward one of the 10 BT09 IRDCs
(IRDC G028.37+00.07, also referred to as IRDC C), including NIR
extinction offset corrections of Kainulainen and Tan (2013), to
produce a higher dynamic range extinction map.

The advantages of such MIREX mapping are high angular resolution
(2$\arcsec$), a near-fully sampled image (except in regions of bright
MIR emission) and an independence to dust temperature (except, again,
in very warm regions that are sources of MIR emission). A major
uncertainty is the need to estimate the background intensity by
interpolation from regions around the cloud. The background intensity
can also have significant small scale fluctuations, which however can
be assessed from the off-cloud regions, but which are an inherent
random uncertainty. Another major issue is the need to estimate the
level of the MIR foreground intensity from the diffuse ISM, since
IRDCs are typically at several kpc distances. The key method to
estimate the foreground is to measure it empirically in spatially
independent ``saturated'' regions of the IRDC where essentially
negligible background light makes it through the cloud (as long as
such regions exist), and then assume it is spatially invariant across
the IRDC or some local region (BT12). The accuracy of measuring the
saturated intensity level depends on the photometric precision of the
image and sets a maximum $\Sigma_{\rm sat}$ that can be probed in the
map.

MIREX mapping does require knowing the $\sim8\:\micron$ opacity per unit
mass: different dust models, such as the thin and thick Ossenkopf and
Henning (1994, hereafter OH94) moderately coagulated models or the
Weingartner \& Draine (2001) bare grain models, show only modest
variation at the level of $\sim 30\%$. Then a (refractory)-dust to gas
mass ratio also needs to be assumed (1:142) is our adopted fiducial
value (Draine 2011; note BT09 and BT12 assumed 1:156). Overall, BT09,
BT12 and BTK14 adopted the OH94 thin ice mantle model ($10^5$~years of
coagulation at density $\sim 10^6\:$cm$^{-3}$) as a fiducial, which
has $\kappa_{\rm 8\mu m}$ (averaged over the IRAC band)
$\simeq$7.5~cm$^2$~g$^{-1}$. With these methods, BT12 could trace
details of highly extincted regions up to, $\Sigma_{\rm
  sat}\sim0.5\:$g~cm$^{-2}$, i.e., $A_V\sim$100~mag. BTK14 could reach
somewhat higher $\Sigma_{\rm sat}\sim1\:$g~cm$^{-2}$, i.e.,
$A_V\sim200$~mag.

LT14 extended MIREX methods to the FIR regime by using {\it
  Spitzer}-MIPS $24\:\micron$ MIPSGAL data (Carey et al. 2009) and
{\it Herschel}-PACS $70\:\micron$ archival data (proposal IDs:
KPGT-okrause-1 \& KPOT-smolinar-1) of IRDC G028.37+00.07. From
studying $\sim1^\prime$ regions around three dense, locally saturated
cores (the white circles in Figure~1), LT14 found tentative evidence
of $\kappa(\lambda)$ variation with $\Sigma$ in the sense that
the relative extinction curves appear to flatten as $\Sigma$
increases. Such behavior is consistent with an evolution from the OH94
thin ice mantle model to the thick ice mantle model, in which there is
a growth of the volume of the ice cover from 0.5$\times$ that of the
refractory component to 4.5$\times$. Some degree of flattening could
also result from coagulation of the grains (Cardelli et al. 1989;
Weingartner \& Draine 2001; Ormel et al. 2011, hereafter O11).

\begin{figure}
\begin{center}$
\begin{array}{ll}
\hspace{0in} \includegraphics[width=3.3in]{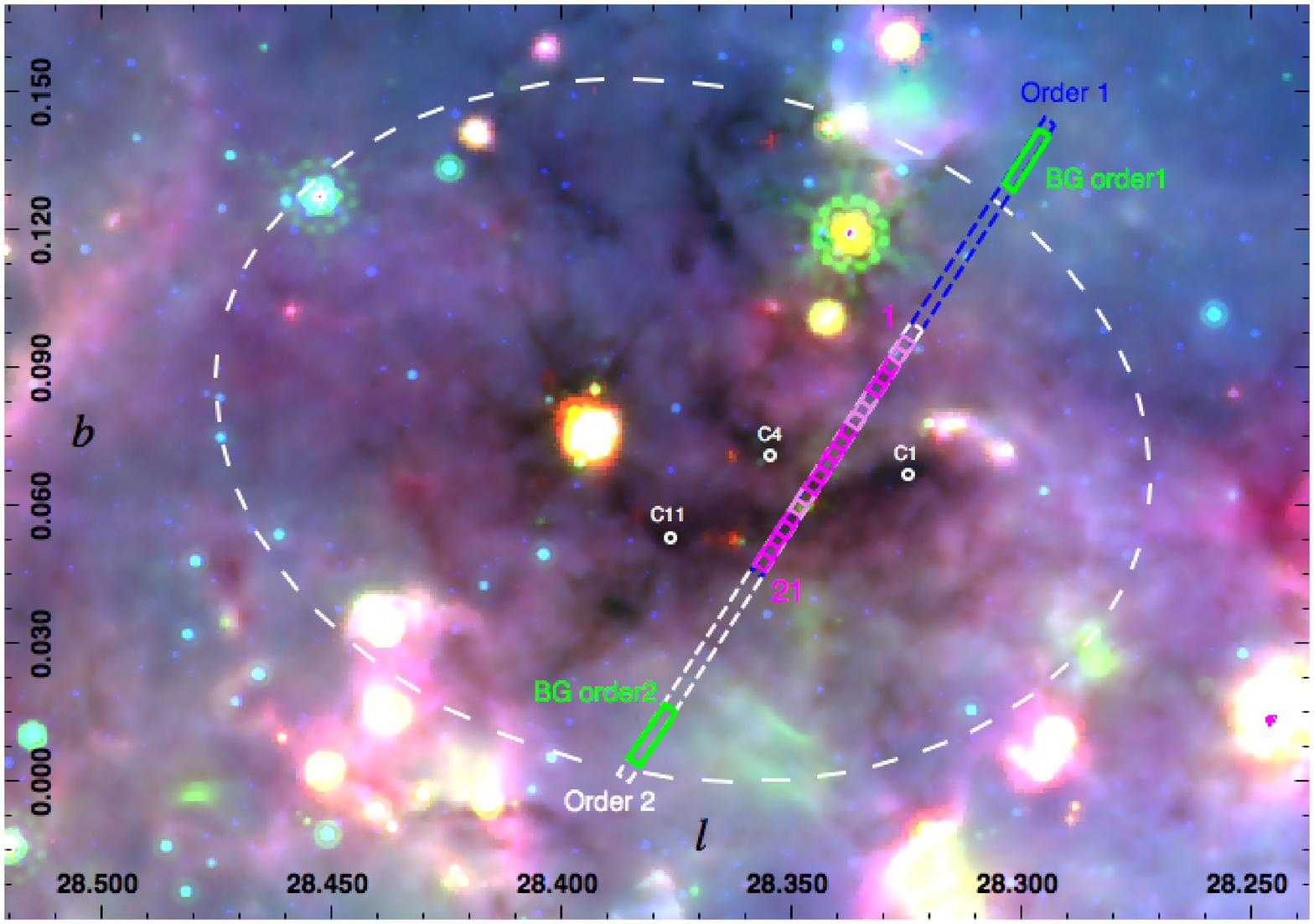}\\
\hspace{0in} \includegraphics[width=3.6135in]{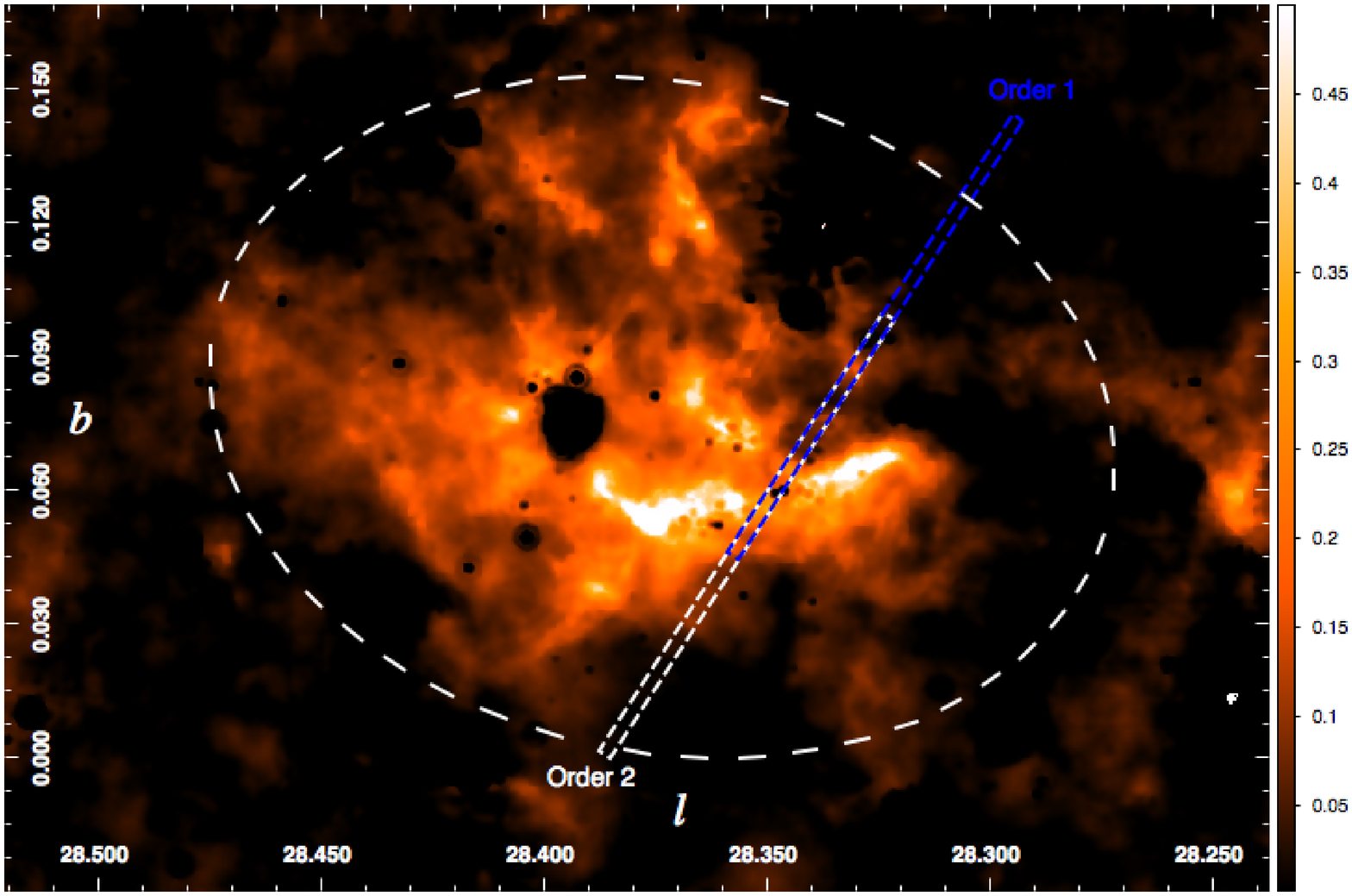}\\

\end{array}$
\end{center}
\caption{\footnotesize 
{\it Upper:} IRDC G028.37+00.07 (Cloud C) as viewed by {\it
  Spitzer}-IRAC $8\:\micron$ (blue), {\it Spitzer}-MIPS $24\:\micron$
(green) and {\it Herschel}-PACS $70\:\micron$ (red). The dashed
ellipse shows an approximate boundary for the IRDC defined by Simon et
al. (2006).  The strip observed by {\it Spitzer}-IRS is shown: Order 1
($19.5$--$38.0\:\micron$) was measured in the white rectangle region;
Order 2 ($14.0$--$21.3\:\micron$) in the blue rectangle. Order 1 and 2
data are both available where these regions overlap, i.e., in the
magenta/violet region, which has been divided into 21 10$\arcsec$
$\times$ 10$\arcsec$ regions (numbered from upper right), from which
individual spectra have been analyzed. Magenta squares indicate
``dark'' regions that are used for our extinction law study. Violet
squares are regions we consider contaminated by $24\:\micron$ point
source emission. The two green regions at the outer ends of the strip
are the regions used to assess the spectrum of the background (BG)
emission from the diffuse ISM. Small white circles indicate the
positions of the dark, saturated cores, C1, C4, C11, that were
investigated by LT14.  {\it Bottom:} Mass surface density map (in $\rm
g\:cm^{-2}$) derived from {\it Spitzer}-MIPS $24\:\micron$ data,
$\Sigma_{\rm 24\mu m}$, of the same region as the RGB map above. The
slit positions of Orders~1 and 2 are shown.}\label{fig:layout}
\end{figure}

 In this paper, we present a new spectroscopic IR extinction (SIREX)
mapping method by using {\it Spitzer}-IRS mid-infrared spectroscopic
data that covers a part of IRDC G028.37+00.07. This IRDC (which we
will also refer to as Cloud C, BT09) is located at 5~kpc distance and
is one of the most massive IRDCs known with $\sim 7\times
10^4\:M_\odot$ within an effective radius of about 8~pc (BTK14) (i.e., the
dashed ellipse shown in Figure~\ref{fig:layout}). It hosts a modest
level of star formation (e.g, the MIR-bright sources in
Fig.~\ref{fig:layout}; see also Zhang et al. 2009),
but overall the star formation activity appears to be relatively low,
and so this cloud should represent an earlier stage of massive star
cluster formation.

The main objective and plan of the paper are as follows. The SIREX
method adopts saturation based MIREX (BT12) and FIREX (LT14) methods
to then allow investigation of the extinction law variation as a
function of mass surface density. In \S2 we review and update the
results of LT14 for extinction law variation in different mass surface
density regions, including utilization of {\it WISE} $12\:\micron$
data. We introduce the detailed methods of SIREX mapping in \S3. In
\S4, we present our results, first demonstrating the measurement of
the Galactic background and foreground from the diffuse ISM, and then
presenting the extinction law in different mass surface density
environments. We discuss the implications of these results and
concluded in \S5.

\begin{figure*}
\begin{center}$
\begin{array}{c}
\hspace{-0.1in} \includegraphics[width=5in]{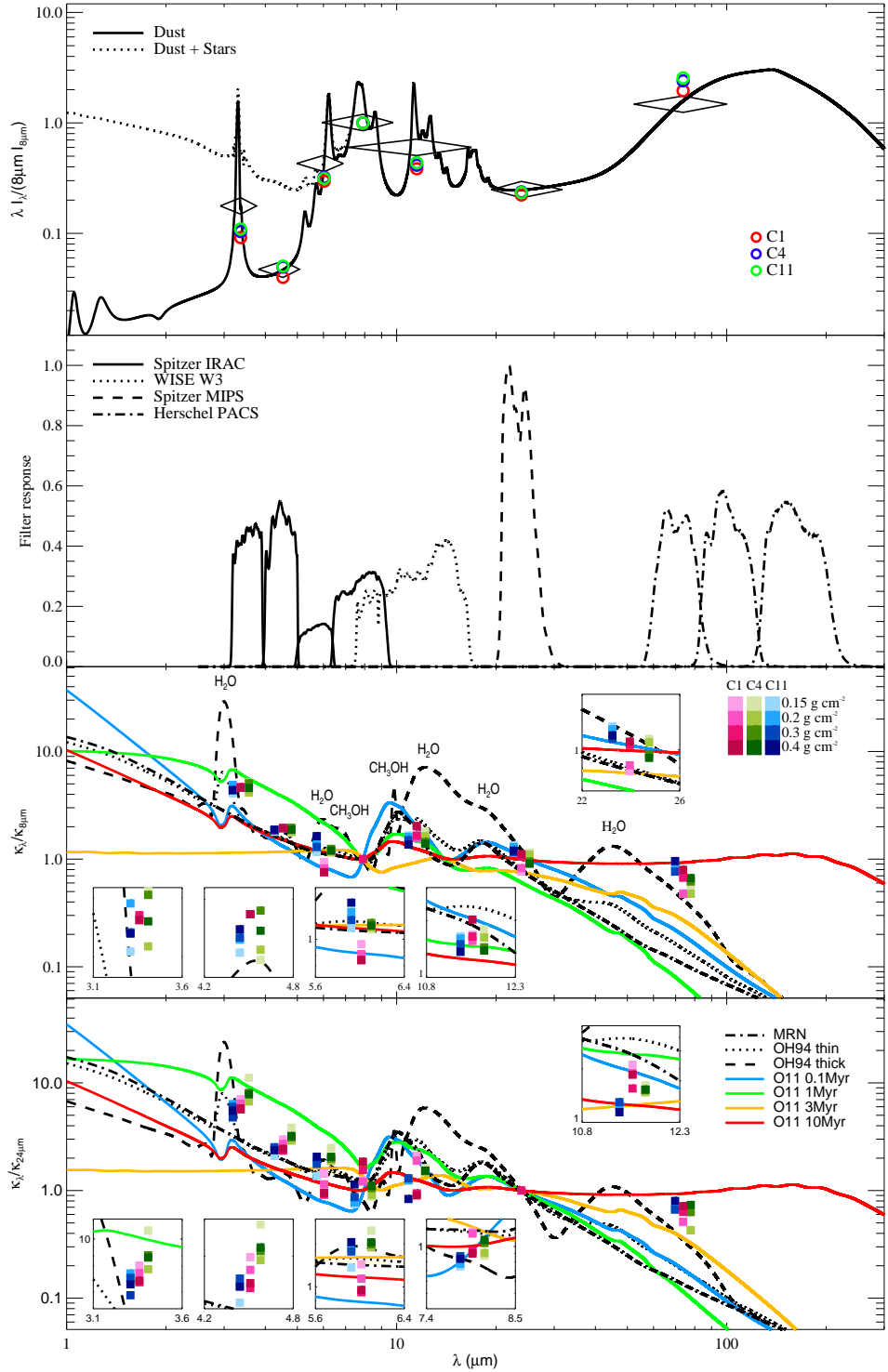}\\
\end{array}$
\end{center}
\caption{
\footnotesize
{\it{(a) Top:}} Total IR to sub-mm SED of the Galactic plane diffuse
ISM, normalized to the {\it Spitzer}-IRAC $8\:\micron$ band flux. The
solid line shows the dust-only component of the DL07 model and the dotted
line indicates the total flux from dust and stellar components.  Large
diamonds indicate convolution of the dust-only SED with the filter
response functions of corresponding instruments (panel b). The red,
blue and green open circles show foreground intensities, relative to
the IRAC $8\:\micron$ value, measured towards the three saturated
cores C1, C4 and C11, respectively.
{\it{(b) Second from top:}} Filter response functions of IRAC bands
1-4 (solid lines), {\it WISE} W3 band at $\sim12\:\micron$ (dotted),
MIPS~$24\:\micron$ (dashed) and PACS~70, 100 \& $160\:\micron$
(dot-dashed).
{\it{(c) Third from top:}} Opacities relative to effective opacity
measured in the IRAC~$8\:\micron$ band. Black dotted/dashed/dot-dashed 
lines show thin/thick ice mantle and bare grain dust models of OH94 
after $10^5$~yr of coagulation at $n_{\rm{H}}=10^6\:{\rm{cm}}^{-3}$ 
(or 10$^6$~yr of coagulation at $n_{\rm{H}}=10^5\:{\rm{cm}}^{-3}$). 
Blue, green, gold and red solid lines indicate O11 fully ice coated 
dust models after 0.1, 1, 3 and 10~Myr of coagulation at
$n_{\rm{H}}=10^5\:{\rm{cm}}^{-3}$, respectively.
The red/green/blue filled squares (for C1/C4/C11 regions; slight
wavelength offsets applied for clarity; see also zoom-in insets) show
$\kappa_{\nu}/\kappa_{\rm 8\mu m}$ for the indicated $\Sigma_{\rm 8\mu m}$ ranges. 
For uncertainties in these data, see Figure~\ref{fig:scatter}.
{\it{(d) Bottom:}} Same as (c), but now showing opacities relative to
effective opacity measured in the MIPS~$24\:\micron$
band. 
}\label{fig:photometry}
\end{figure*}

\section{Photometric analysis}\label{S:photometry}

Here we update the photometric MIREX and FIREX analysis methods of
LT14 as applied to IRDC C, including: utilizing the improved
$8\:\micron$ extinction map of BTK14, which probes to higher values of
$\Sigma$; comparing to the DL07 diffuse ISM dust emission model;
analyzing {\it WISE} $12\:\micron$ data; and comparing to O11 dust
opacity models, in addition to those of OH94.

\subsection{Dust Model Opacities in MIR to FIR Photometric Bands}

Following the methods of LT14, we consider several different dust
models from OH94 and now also O11. The thin \& thick ice mantle models
of OH94 (ice composition of H$_2$O:CH$_3$OH:CO:NH$_3$=100:10:1:1) and
the models of O11 (ice composition of
H$_2$O:CH$_3$OH:CO:NH$_3$=100:66:5:5) are moderately coagulated grains
(i.e., coagulation for various periods at $n_{\rm H}=10^6\:{\rm
  cm}^{-3}$ for OH94 and at $n_{\rm H}=10^5\:{\rm cm}^{-3}$ for O11.
For consistency with LT14, we utilized 10$^5$~yr of coagulation OH94
models (which is equivalent to 10$^6$~yr of coagulation at $n_{\rm
  H}=10^5\:{\rm cm}^{-3}$). We also adopt the gas to (refractory
component) dust mass ratio of 142 (Draine 2011) that is used in LT14.

 Densities of $n_{\rm H}\sim 10^5\:{\rm cm^{-3}}$ are typical of dense
cores and clumps within IRDCs (e.g., Tan et al. 2014), however the
lifetimes of these structures are less well constrained. We expect the
clouds to be at least one local free-fall time old, i.e., $1.4\times
10^5\:{\rm yr}$ at this density, but ages that are $\sim10\times$
longer are certainly a possibility and may help account for high
levels of deuteration that are seen in some IRDCs (Kong et al. 2015).
The longest duration O11 model involves coagulation for $10^7\:{\rm
  yr}$. The applicability of such timescales to dense clumps and cores
seems less likely to be valid, especially given the apparent turbulent
nature of these molecular clouds, which should induce density
fluctuations on shorter timescales. We note that the OH94 and O11 are
local models, valid for particular, uniform densities (c.f., the
structured core models of Weidenschilling \& Ruzmaikina 1994).
We note also that the OH94 and O11 models involve dust grain
coagulation that is mediated by relative grain velocities that are set
by a model of subsonic turbulence. On the larger clump scales within
IRDCs, we expect supersonic turbulent motions to be more relevant,
although these may be moderated by the presence of strong magnetic
fields (Pillai et al. 2015).
Supersonic turbulence is seen to decay within a few free-fall times in
numerical simulations (Stone et al. 1998; Mac Low et al. 1998; see
review by McKee \& Ostriker 2007), 
so on the smaller scales of dense
cores we may expect conditions to have evolved closer to a subsonic
turbulent cascade.

We derive filter and background spectrum weighted opacity values. BT09
derived these for IRAC bands 1-4 and the MIPS $24\:\micron$ band using
the diffuse Galactic background spectrum from Li \& Draine (2001; 
hereafter LD01) and a total to refractory dust component mass ratio of 
156. LT14 modified these values based on the total to refractory component 
dust mass ratio of 142 and expanded to {\it Herschel}-PACS 70, 100 and
$160\:\micron$ filters. Here, we recalculate these opacities by
utilizing the diffuse Galactic background emission model of DL07, and
now also including the {\it WISE} $12\:\micron$ band filter
(Table~\ref{tb:kappa}).

These opacities can be used to calculate $\Sigma$ from a given set of
imaging data. Their relative values will also be compared to those
observed in IRDC C.

\begin{deluxetable*}{lccccccccc}
\tabletypesize{\footnotesize}
\tablecolumns{10}
\tablewidth{0pt}
\tablecaption{Telescope Band and Background-Weighted Dust Opacities Per Total Mass\tablenotemark{a} ($\rm{cm^{2}\:g^{-1}}$)}
\tablehead{\colhead{Dust Model\tablenotemark{b}}                                              &
           \colhead{IRAC3.5} &  
           \colhead{IRAC4.5} &  
           \colhead{IRAC6} &  
           \colhead{IRAC8} &
           \colhead{WISE12} &
           \colhead{MIPS24} &
           \colhead{PACS70} &
           \colhead{PACS100} &
           \colhead{PACS160}\\
	   \colhead{} &
	   \colhead{$\rm{3.52\:\micron}$\tablenotemark{c}} &
	   \colhead{$\rm{4.49\:\micron}$} &
	   \colhead{$\rm{5.91\:\micron}$} &
	   \colhead{$\rm{7.80\:\micron}$} &
	   \colhead{$\rm{12.0\:\micron}$} &
	   \colhead{$\rm{23.0\:\micron}$} &
	   \colhead{$\rm{74.0\:\micron}$} &
	   \colhead{$\rm{103.6\:\micron}$} &
	   \colhead{$\rm{161.6\:\micron}$} 
}
\startdata
OH94 thin mantle 	& 26.9 (19.5) & 14.4 (11.7) & 10.31 (9.37) & 8.19 (8.04) & 12.8 & 6.26 & 1.14 & 0.598 & 0.287\\
OH94 thick mantle 	& 52.4 (38.0) & 17.5 (14.3) & 20.1 (18.2) & 10.3 (10.1) & 36.7 & 11.4 & 3.96 & 1.26 & 0.399\\
O11 30ky & 12.3 (11.9)   & 7.03 (6.91) & 4.79 (4.75) & 7.14 (7.12) & 8.24 & 5.01 & 0.79 & 0.38 & 0.16\\
O11 0.1My	& 16.3 (14.1) & 8.16 (7.57) & 5.14 (4.97) & 7.60 (7.56) & 9.87 & 6.59 & 1.22 & 0.590 & 0.242\\
O11 0.3My	& 41.4 (25.7) & 18.4 (12.2) & 9.08 (6.64) & 8.91 (8.13) &  11.5 (11.0) & 7.34 (7.29) & 1.25 & 0.59 & 0.24\\
O11 1My	& 115 (61.3) & 70.5 (37.8) & 40.3 (22.0) & 21.7 (14.1) & 22.5 (16.0) & 11.4 (9.17) & 1.34 (1.19) & 0.57 (0.53) & 0.21 (0.20)\\
O11 3My	& 30.9 (17.9) & 31.5 (17.7) & 32.4 (17.8) & 26.9 (15.7) & 25.4 (15.7) & 20.1 (12.8) & 7.30 (4.36) & 3.49 (2.09) & 1.13 (0.70)\\
O11 10My	& 10.8 (6.65) & 7.22 (4.55) & 5.53 (3.57) & 5.05 (3.59) & 5.41 (4.02) & 4.74 (3.49) & 4.45 (2.89) & 4.84 (2.93) & 5.14 (2.87)
\enddata
\tablenotetext{a}{
A total to refractory dust mass ratio of 142 is adopted (Draine 2011).
 OH94 opacities in the IRAC bands have been scaled from values in
  parentheses to include an approximate estimate of the contribution
  from scattering, based on the Weingartner \& Draine (2001) dust
  model for $R_V=5.5$ dust. The contribution of scattering to the
  total extinction is expected to be very small for $\lambda>10\:{\rm
    \mu m}$ for these dust models. The online O11 models provide
  scattering opacities, so we compute both the pure absorption
  extinction (in parentheses; shown when there is a significant
  difference) and total extinction including scattering.
}
\tablenotetext{b}{
OH94 models for 10$^5$~yr of coagulation at $n_{\rm H}=10^6\:{\rm
  cm}^{-3}$ (equivalent to 10$^6$~yr of coagulation at $n_{\rm
  H}=10^5\:{\rm cm}^{-3}$). Fully ice covered O11 models for various
coagulation times as noted (at $n_{\rm H}=10^5\:{\rm cm}^{-3}$).}
\tablenotetext{c}{Mean wavelengths weighted by filter response and
  background spectrum.}
\label{tb:kappa}
\end{deluxetable*}

\begin{figure*}
\begin{center}$
\begin{array}{c}
\hspace{-0.1in} \includegraphics[width=5.5in]{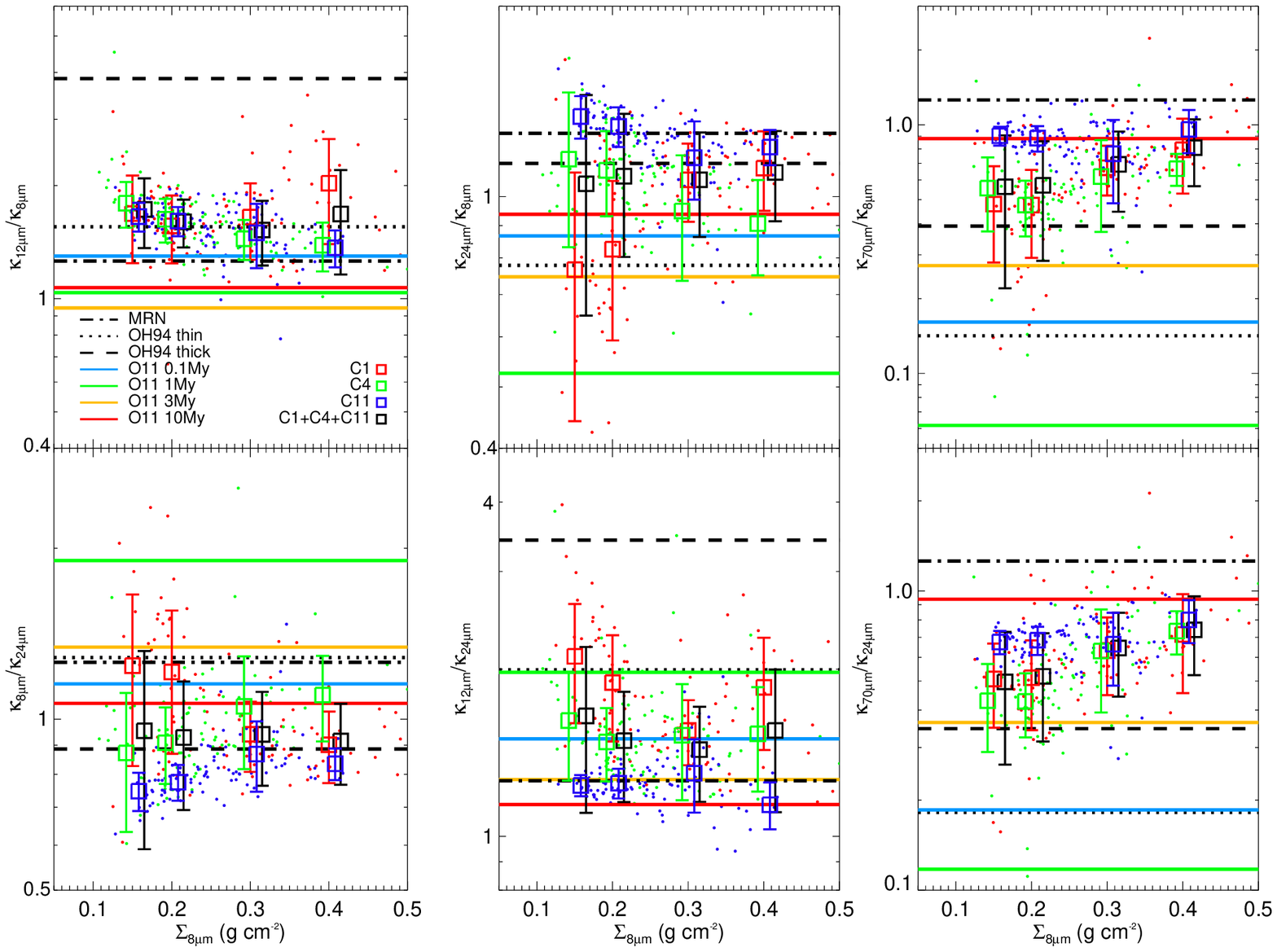}\\
\end{array}$
\end{center}
\caption{\footnotesize 
Scatter plots of $\kappa_{\rm 12\mu m}/\kappa_{\rm 8\mu m}$,
$\kappa_{\rm 24\mu m}/\kappa_{\rm 8\mu m}$, $\kappa_{\rm 70\mu
  m}/\kappa_{\rm 8\mu m}$, $\kappa_{\rm 8\mu m}/\kappa_{\rm 24\mu m}$,
$\kappa_{\rm 12\mu m}/\kappa_{\rm 24\mu m}$ and $\kappa_{\rm 70\mu
  m}/\kappa_{\rm 24\mu m}$ versus $\Sigma_{\rm 8\mu m}$, as measured
from the BTK14 $\Sigma_{\rm 8\mu m}$ map and imaging data from
{\it Spitzer}-IRAC $8\:\micron$, {\it WISE} $12\:\micron$, {\it
  Spitzer}-MIPS $24\:\micron$ and {\it Herschel}-PACS
$70\:\micron$. The size of pixels in all maps have been regridded to
$\sim6\arcsec$ resolution (i.e., the approximate beam sizes of the
$12$, $24$, and $70\:\micron$ images). Red, green and blue dots
correspond to pixels in C1, C4 and C11 regions, respectively. The open
squares are the mean values of log$_{10}$ of relative $\kappa$, with
error bars indicating the $1\sigma$ dispersion.
The black open squares are the overall average of the three cores.
Black dotted, dashed and dot-dashed lines are relative $\kappa$ values of OH94
thin, thick ice mantle and bare grain models, respectively.  Blue, green, gold and
red solid lines indicate O11 fully ice coated models of 0.1, 1, 3 and
10~Myr of coagulation time, respectively.}
\label{fig:scatter}
\end{figure*}

\subsection{The MIR to FIR Dust Extinction Law in IRDCs from Broad Band Photometry}

We now re-visit the tentative evidence presented by LT14 of extinction
law variation, i.e., flatter $\kappa_{\lambda}/\kappa_{\rm 8\mu m}$, in
higher $\Sigma$ regions of IRDC C. We utilize the same MIR and FIR
images processed by LT14, i.e., {\it Spitzer}-IRAC 3.5, 4.5, 5.9,
$8\:\micron$, {\it Spitzer}-MIPS $24\:\micron$ and archival data of
{\it Herschel}-PACS $70\:\micron$. The angular resolutions of the IRAC
bands, MIPS $24\:\micron$ band and PACS $70\:\micron$ band are
$\sim$2$\arcsec$, 6$\arcsec$ and 6$\arcsec$, respectively.

We also calibrate and use {\it WISE} $12\:\micron$ imaging data
with angular resolution $\sim$6.5$\arcsec$. Following MIREX (BT09,
BT12) and FIREX mapping (LT14) methods, we estimated the background
emission by applying the small median filter (SMF) method, i.e., a
median filter size set to 1/3 of the major axis of the IRDC, i.e.,
4\arcmin, in the region outside the dashed ellipse shown in
Fig.~\ref{fig:layout}, followed by interpolation inside the ellipse.

We also adopted the saturation based foreground estimation method of
BT12: the values of the foreground, relative to that in IRAC band 4 at
$8\:\micron$, were measured towards cores C1, C4 and C11 and are shown
in the top panel of Fig~\ref{fig:photometry}. They are compared to the
DL07 model of Galactic plane diffuse ISM emission (excluding stars),
which is shown by the solid line, with the diamond shaped boxes
showing convolution of this model with the instrumental filter
response functions.

The $\Sigma_{\rm 8\mu m}$ map of IRDC C from BTK14, which probes to
higher mass surface densities than the BT12 map, is now utilized so
that we are able to consider binned values up to 0.4~g~cm$^{-2}$,
compared with the 0.3~g~cm$^{-2}$ of LT14. Following LT14, we derive
optical depth maps in 1\arcmin\ regions around each saturated core C1,
C4 and C11 in each waveband and then maps of relative opacity,
normalized to that of IRAC band 4 at $8\:\micron$, and with angular
resolution of $\sim6\arcsec$, set by the PACS~$70\:\micron$ image. We
also consider opacities normalized to those derived from MIPS images
at $24\:\micron$.

 Note, that this analysis, following BT12 and LT14, assumes that the
IRDC itself is a negligible source of emission, either intrinsically
(e.g., from thermal emission of surface layers of warm dust, including
transiently heated small grains) or from scattering of radiation into
our line of sight from other directions by dust within the IRDC. This
assumption appears to be supported by the fact that independent dark
saturated regions are seen within the IRDC across the studied MIR to
FIR wavebands. If scattering has an effect, it is expected to be most
important at the shorter, MIR wavelengths, where it has been seen to
occur in deep IRAC 3.5~$\rm mu m$ and 4.5~$\rm \mu m$ images of nearby
cores, producing the ``core shine'' phenomenon (e.g., Pagani et
al. 2010; Lef\`evre et al. 2014).
Accurate assessment of the effects of such scattering in IRDCs
requires radiative transfer modeling and assumptions about the
radiation fields and 3D IRDC structure. Given our focus in ths paper
on dust extinction properties at longer wavelengths, especially
extending from $\sim 10\:{\rm \mu m}$ to $\sim 70\:{\rm \mu m}$, where
scattering should be of relatively minor importance for most grain
models, we continue with the approximation adopted in our previous
studies of neglecting the effects scattering.

Each pixel in the maps has a value of $\Sigma_{\rm 8\mu m}$. Scatter
plots of $\kappa_{\rm 12\mu m}/\kappa_{\rm 8\mu m}$, $\kappa_{\rm
  24\mu m}/\kappa_{\rm 8\mu m}$ and $\kappa_{\rm 70\mu m}/\kappa_{\rm
  8\mu m}$ versus $\Sigma_{\rm 8\mu m}$ are shown in the top row of
Fig.~\ref{fig:scatter}. The bottom row shows the equivalent plots of
$\kappa_{\rm 8\mu m}/\kappa_{\rm 24\mu m}$, $\kappa_{\rm 12\mu
  m}/\kappa_{\rm 24\mu m}$ and $\kappa_{\rm 70\mu m}/\kappa_{\rm 24\mu
  m}$ versus $\Sigma_{\rm 8\mu m}$. Note, with the new BTK14 $\Sigma$
map, including its NIR offset corrections, there are relatively few
pixels with $\Sigma<0.1\:$g~cm$^{-2}$. Thus, we replace the
low-$\Sigma$ bin of $0.1\pm0.05\:$g~cm$^{-2}$ used by LT14, with a bin
defined by $\Sigma=0.15\pm0.05\:$g~cm$^{-2}$. Binned averages for
$\Sigma_{\rm 8\mu m}= 0.15\pm0.05, 0.2\pm0.05, 0.3\pm0.05,
0.4\pm0.05$~g~cm$^{-2}$ are shown -- these are values of $\Sigma$ that
should not be affected by saturation in the BTK14 map (these values
are also shown in the third and fourth panels of
Fig.~\ref{fig:photometry}). Higher values of binned
$\Sigma_{8\micron}$ are shown in Fig.~\ref{fig:scatter}, but these
results should be treated with caution since they may be affected by
saturation, especially affecting both 8 and 24~$\rm \mu m$ results.

Comparing to the models of OH94, the observed $\kappa_{\rm 70\mu
  m}/\kappa_{\rm 8\mu m}$ and $\kappa_{\rm 70\mu m}/\kappa_{\rm 24\mu
  m}$ values are closer to the thick ice mantle case than the thin ice
mantle case (similar to the findings of LT14). However, the 12~$\rm
\mu m$ {\it WISE} results do not favor the thick ice mantle model (see
also Fig.~\ref{fig:photometry}. Comparing to the O11 models, the
observed $\kappa_{\rm 70\mu m}/\kappa_{\rm 8\mu m}$ and $\kappa_{\rm
  70\mu m}/\kappa_{\rm 24\mu m}$ values are consistent with models
with coagulation times of $\gtrsim$3~Myr. However, again the shorter
wavelength results (including the IRAC bands 1 to 3; see
Fig.~\ref{fig:photometry}) tend to favor the shorter coagulation time
models. In other words, there is not a single dust evolution model
that provides a fully consistent fit to the data.

We next examine whether there is any trend of evolving opacity law
with $\Sigma$. Inspecting the panels of Fig.~\ref{fig:scatter}, we
notice that the $\kappa_{\rm 70\mu m}/\kappa_{\rm 8\mu m}$ and
$\kappa_{\rm 70\mu m}/\kappa_{\rm 24\mu m}$ values appear to increase
as $\Sigma$ increases. We examine the Pearson correlation
coefficients, $r$, and $p$-values (probability of chance correlation)
for these cases finding $r=0.25,0.34$ and
$p=5\times10^{-5},1\times10^{-8}$ when analysing all the data for
$\kappa_{\rm 70\mu m}/\kappa_{\rm 8\mu m}$ and $\kappa_{\rm 70\mu
  m}/\kappa_{\rm 24\mu m}$, respectively. These results appear to
indicate evidence for a systematically evolving MIR to FIR extinction
law, i.e., becoming flatter, as mass surface densities (and presumably
also volume densities) increase. Such a flattening is expected from
models of grain growth (e.g., the O11 models). The other wavelength
data do not show trends with $\Sigma$ that are as
significant. However, this may be expected since, e.g., the O11 models
also do not predict as significant a variation as they do in
comparison with the evolution of $\kappa_{\rm 70\mu m}$.

We conclude that the 70~$\rm \mu m$ data in comparison with the 8 and
24~$\rm \mu m$ data show evidence for growth of dust grains via
coagulation (O11) and/or thickening of ice mantles (OH94, although
such models struggle to match shorter wavelength results) in the
densest regions. However, the particular shape of the extinction law
and its possible variation are only quite poorly constrained.

\section{Spectroscopic IR Extinction (SIREX) Mapping Methods}\label{S:sirexmethod}

\subsection{MIR Spectroscopic data of IRDC G028.37+00.07}

\begin{figure}
\begin{center}$
\begin{array}{c}
\hspace{0in} \includegraphics[width=3.4in]{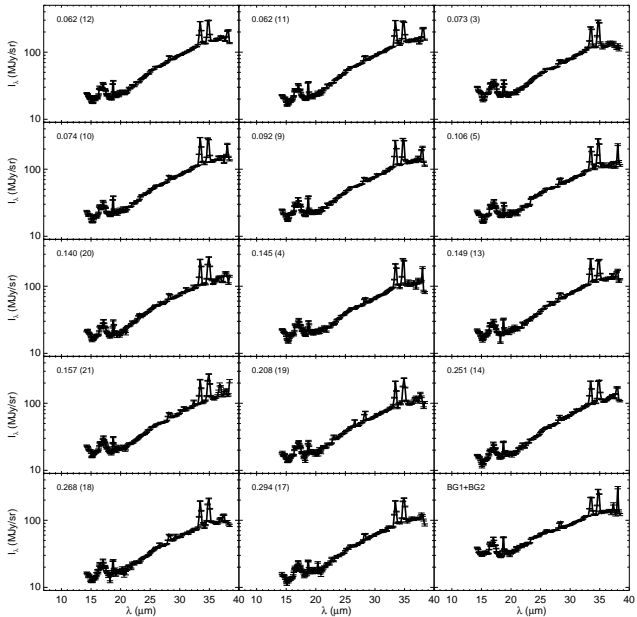}\\
\end{array}$
\end{center}
\caption{
\footnotesize Extracted spectra of the selected 14 ``dark'' regions
(i.e., without contamination of point sources) arranged in order of
increasing $\Sigma_{\rm 24\mu m}$ (shown in top left corner of each panel
in g~cm$^{-2}$; the number in parentheses is the region number, see
Fig.~1). The summed spectrum from the background (BG) regions is shown
in the lower-right panel. The photometric error bars are estimated by
CUBISM. All spectra are shown before the zodiacal correction has been applied.
}
\label{fig:flucori}
\end{figure}

We analyze archival {\it Spitzer}-IRS Long-Low spectra (program:
3121, PI: Kraemer, K. E.) with spectral resolution $R\sim$57--126 of
IRDC G028.37+00.07 (Cloud C). The IRS Long-Low module consists of two
long slits (1st Order and 2nd Order; hereafter LL1 and LL2) that have
a size of 168$\arcsec\times$10.5$\arcsec$ each.
LL1 and LL2 observe the wavelength ranges of $19.5$--$38.0\:\micron$
and $14$--$21.3\:\micron$, respectively. The IRS spectral data of IRDC
C include 4 Long-Low observations of staring modes that cover
$\sim7\arcmin$ in length for each order (LL1 and LL2) diagonally
across the IRDC (Fig.~\ref{fig:layout}). The overlapped region of LL1
and LL2 is $\sim4\arcmin$ long. The extraction of the observed spectra
is executed by utilizing the CUbe Builder for IRS Spectra Maps
(CUBISM, Smith et al. 2007). The extracted spectra before any
corrections are shown in Figure~\ref{fig:flucori}.

\subsection{Radiative Transfer Assumptions}

We follow the same simple one-dimensional radiative transfer model of
the MIREX (BT12) and FIREX (LT14) methods and here review the basic
equations that describe the input quantities that are required for the
analysis.  We need to know the intensity of radiation directed toward
the observer at the location just behind, $I_{\nu,0}$, and just in
front, $I_{\nu,1}$, of the target IRDC. The infrared emission from the
IRDC (both intrinsic emission and that scattered into the line of
sight) is assumed to be negligible so that
\begin{equation}
I_{\nu,1}=e^{-\tau_{\nu}}I_{\nu,0},
\end{equation}
where optical depth $\tau_{\nu}=\kappa_{\nu}\Sigma$, $\kappa_{\nu}$ is
total (absorption $+$ scattering) opacity at frequency $\nu$ per unit total mass and $\Sigma$ is
the total mass surface density. The spectrum of $I_{\nu,0}$, i.e., the
background, is to be estimated from calibrated spectra of
``off-positions'' away from the target IRDC region, while $I_{\nu,1}$
is derived from the observed spectrum toward the location of interest.

However, we cannot observe $I_{\nu,0}$ and $I_{\nu,1}$ directly
because these intensities are contaminated by foreground emission,
$I_{\rm \nu,fore}$ so that the observed background emission, 
$I_{\rm \nu,0,obs}$, and the observed intensity towards the cloud, 
$I_{\rm \nu,1,obs}$, are
\begin{equation}
I_{\rm \nu,0,obs}=I_{\nu,0}+I_{\rm \nu,fore}
\label{eq:sat1}
\end{equation} 
and 
\begin{equation}
I_{\rm \nu,1,obs}=I_{\nu,1}+I_{\rm \nu,fore},
\label{eq:sat2}
\end{equation}
respectively. Here we are assuming that $I_{\rm \nu,fore}$, which is
emitted from the diffuse ISM, does not vary significantly in its
intensity over the face of the IRDC and to the off-position.

Therefore, as with MIREX and FIREX mapping, SIREX mapping also
requires knowing the intensity of the foreground emission toward the
target region. While the MIREX and FIREX derived $I_{\rm \nu,fore}$
from observed ``saturated'' regions, i.e., they have same observed
intensity that agrees with the minimum intensity within a range set by
2$\sigma$ (or other set range) noise level of the images, SIREX
mapping toward IRDC C cannot apply this same saturation-based
foreground method, since there is no slit position that is coincident
with a saturated core, i.e., as defined from the MIPS $24\:\micron$
maps (LT14) (see Fig.~1). Thus we need to apply another method to
estimate the foreground spectrum.

\subsection{Background and Foreground Estimation}

We select two background regions, i.e., off-positions, from near the
north (BG1) and south (BG2) ends of the slit position
($\sim$50$\arcsec\times$10$\arcsec$ on each side). Unfortunately,
these regions are not covered by both orders
(Fig.~\ref{fig:layout}). The selected background regions have a
$\sim$12\% difference in their average {\it Spitzer}-MIPS
$24\:\micron$ photometric intensities, $\la$5\% differences in {\it
  Spitzer}-IRAC $8\:\micron$, $\sim$10\% differences {\it WISE}
$12\:\micron$ intensities and $\sim$35\% differences {\it
  Herschel}-PACS $70\:\micron$. These gradients are utilized, below,
for linear interpolation of the background spectrum along the slit.

\begin{figure}
\begin{center}$
\begin{array}{c}
\hspace{0in} \includegraphics[width=3.4in]{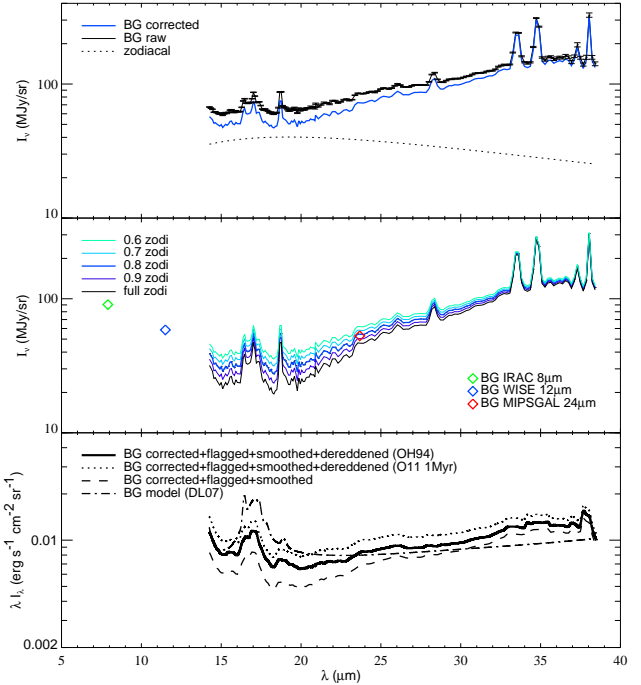}\\
\end{array}$
\end{center}
\caption{
\footnotesize {\it (a) Top:} {\it Spitzer}-IRS spectrum from the
background (BG) regions (green rectangles in Fig.~1a). The black solid
line shows the originally extracted spectrum without any correction
(hereafter 
BG raw) with photometric errors. The black dotted line is a model
spectrum of $\sim$0.8$\times$ the fiducial zodiacal light model
(hereafter zodi). The blue solid line is the
background spectrum after being corrected, i.e., BG raw -
0.785$\times$ zodi.
{\it (b) Middle:} BG region spectra showing the effect of different
fractions of zodiacal light subtraction as a correction to the
background spectrum. The constraint of matching the {\it Spitzer}-MIPS
$24\:\micron$ filter-weighted data point, 
i.e., the average of the
small median filter background model of LT14 of the two BG regions
(red diamond) and also of 21 regions, sets our choice of using 
0.785$\times$ the fiducial zodi model intensity.
The green and blue diamonds indicate similar {\it Spitzer}-IRAC
$8\:\micron$ and {\it WISE} $12\:\micron$ photometric data,
respectively, for the BG regions. {\it (c) Bottom:} Spectral line
masked (see text), smoothed (to $\sim0.5\:\micron$ resolution at
$\sim24\:\micron$) and zodi-corrected background spectrum (dashed
line), and then 
iteratively 
de-reddened via OH94 thin ice mantle model (see text)
background spectrum (solid line).
If the O11 1~Myr dust model is used for de-reddening the background then the 
result is the dotted line, i.e., showing about 10\%--20\% differences
depending on wavelength.
The dot-dashed line indicates the DL07 model of the diffuse Galactic
background emission.}
\label{fig:zodi}
\end{figure}

We compare the {\it Spitzer}-IRS spectroscopic data with the {\it
  Spitzer}-MIPS $24\:\micron$ image of the background and selected 21
regions.  We subtract fiducial model zodiacal emission from the
extracted spectra to obtain intrinsic intensities as a function of
wavelength. However, we find that the intensities of the MIPS
$24\:\micron$ filter-weighted IRS background spectrum after these
zodiacal light corrections have $\sim$10\% higher values than in the
MIPSGAL image. In order to correct this offset, we consider
different fractions of the fiducial model zodiacal light subtraction
from the spectra of the background and the 21 selected regions
(Fig.~\ref{fig:zodi}, middle panel). We find that the average value
of the best matched fraction of the zodiacal model from all regions is
78.5$\%$.

\begin{figure}
\begin{center}$
\begin{array}{c}
\hspace{0in} \includegraphics[width=3.4in]{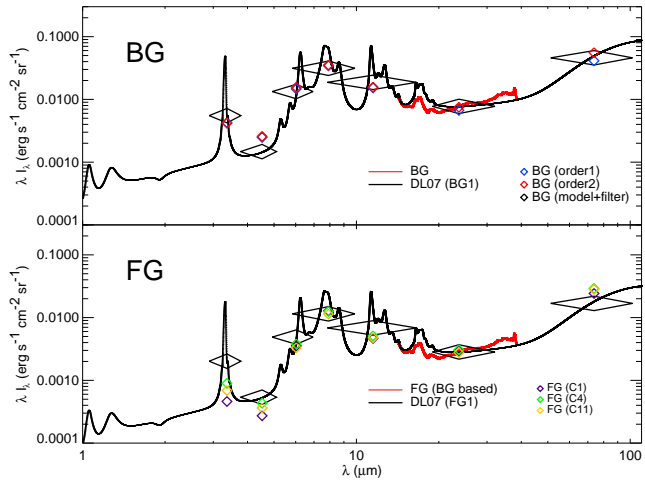}\\
\end{array}$
\end{center}
\caption{
\footnotesize SED of the diffuse Galactic ISM. {\it (a) Top:} Derived
SED of the background emission, $I_{\nu,0}$, from the BG regions. The
red solid line shows the de-reddened 
and foreground-subtracted {\it Spitzer}-IRS data (see text). The small
diamonds show photometric estimates from the BG regions (based on SMF
background models). The black solid line shows the DL07 model (dust
only; best-fit normalized to the {\it Spitzer}-IRS data) and the large
diamonds show its convolution with the filter response functions of
the {\it Spitzer}-IRAC bands, {\it WISE} $12\:\micron$, {\it
  Spitzer}-MIPS $24\:\micron$ and {\it Herschel}-PACS $70\:\micron$
bands. {\it (b) Bottom:} As for (a) but now showing derived SED of the
foreground emission towards IRDC C. Now the small diamonds show
photometric measurements towards the three saturated cores C1, C4 and
C11.
}
\label{fig:bgfgcomp1}
\end{figure}

While the BG1 and BG2 regions are near the ends of the IRS long-slit
aperture, they are still quite close to the IRDC and are thus likely
to be affected by extinction from the IRDC and its surrounding
GMC. Thus we de-redden the BG spectra to obtain the intrinsic
background spectrum that is not contaminated by the IRDC. We first
flag and mask the molecular emission lines (S(III):18.713~$\micron$,
H$_2$ S(0):~28.221~$\micron$, S(III):~33.480~$\micron$ and
Si(II):~34.815~$\micron$) to simplify the spectrum. Then, de-reddening
is applied to estimate $I_{\nu,0}$, via $I_{\nu,0} = I_{\nu,1}
e^{\tau_{\nu}} = (I_{\rm \nu,1,obs}-I_{\rm \nu,fore}) e^{\tau_{\nu}}$
(Fig.~\ref{fig:zodi}c).

To estimate the amount of de-reddening, we adopt the $\Sigma_{\rm
  24\mu m}$ map from the work of LT14 (see Fig. 1) and $\kappa_{\nu}$
from the OH94 thin ice mantle model. Note, the validity of this
assumption can be checked after the fact by comparing our derived BG
spectrum with the DL07 (empirically-calibrated) model, but, still,
this is a fundamental limitation of the SIREX method when the off
positions are too close to the IRDC, which ultimately introduces a
systematic uncertainty in the shape of the derived IRDC extinction
curves, although not affecting their relative shapes.
To test the sensitivity of our results to this choice, we also try
de-reddening the background emission using the O11 model of 1~Myr
coagulation time. 
This leads to a $\sim20\%$ higher intensity of the derived background
(at 24~$\rm \mu m$) than when using the OH94 thin ice mantle model
(Figure~\ref{fig:zodi}).

Having estimated the background spectrum, we utilize it to estimate
the foreground spectrum by scaling the background by a factor,
$\simeq$0.35, derived by considering the ratio of the average MIPSGAL
$24\:\micron$ intensity towards the three saturated cores C1, C4 and
C11 compared with the average MIPSGAL $24\:\micron$ SMF background map
intensity of the BG1 and BG2 regions.

This estimated foreground is then used, iteratively, in the above
de-reddening analysis, since the foreground is not extincted by the
IRDC.  We find this iterative method converges quickly, yielding our
final estimated background spectrum ($I_{\nu,0}$) from the BG regions
and scaled foreground spectrum ($I_{\rm \nu,fore}$). These are shown
in Figure \ref{fig:bgfgcomp1}, together with our photometric estimates
that probe these spectra. We also compare to the best-fitting DL07
model.

\begin{figure}
\begin{center}$
\begin{array}{c}
\hspace{0in} \includegraphics[width=3.4in]{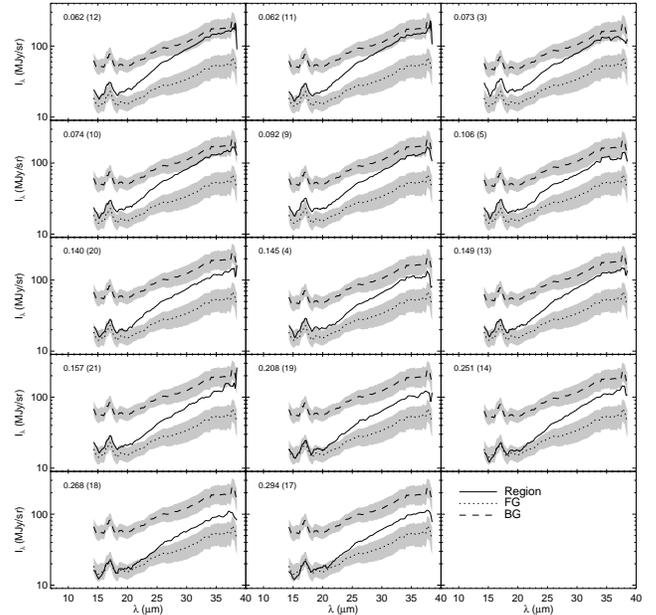}\\
\end{array}$
\end{center}
\caption{
\footnotesize Derived background spectrum (dashed line; grey shaded
region shows estimated errors -- see text), foreground spectrum
(scaled up by a factor $\sim$0.35 from the background), and zodi-corrected
observed spectra (solid lines) from each of the 14 $24\:\micron$-dark
regions (separate panels, in order of increasing $\Sigma_{\rm 24\mu m}$
(indicated in g cm$^{-2}$ in top left corners, along with region
numbers in parentheses).
}
\label{fig:bgfluc1}
\end{figure}

After obtaining an estimate of the background spectrum from the BG
regions, we then consider how it may vary spatially along the IRS
strip. To do this we utilize {\it WISE} 12~${\rm \mu m}$, {\it
  Spitzer}-MIPS 24~${\rm \mu m}$ and {\it Herschel}-PACS 70~${\rm \mu
  m}$ photometric estimates of the background intensities (via the SMF
method) at each end of the slit. Spatial gradients at these
wavelengths are measured: at 12 and 24~${\rm \mu m}$ the intensities
vary by about 10\% across the length of the slit; at 70~${\rm \mu m}$
the variation is about 35\%. We use these results to carry out linear
interpolation of the background intensity as a function of position
and wavelength. The estimated background, i.e., $I_{\nu,0,{\rm obs}}$,
is then compared to the observed emission, i.e., $I_{\nu,1,{\rm
    obs}}$, from each of the 21 dark regions along the slit length. We
find that an overall scaling of a 15\% increase in the background
intensity is necessary to achieve self-consistent opacity results
(see \S\ref{S:results}). This correction is within the estimated size
of systematic errors associated with the above methods, and may be due
to some combination of errors in the background de-reddening,
background spatial interpolation, foreground fraction estimation and
zodiacal correction methods.  Our final adopted estimates for
$I_{\nu,0,{\rm obs}}$ and $I_{\nu,1,{\rm obs}}$ are shown for each
region in Figure~\ref{fig:bgfluc1}.

\begin{figure}
\begin{center}$
\begin{array}{c}
\hspace{0in} \includegraphics[width=3.4in]{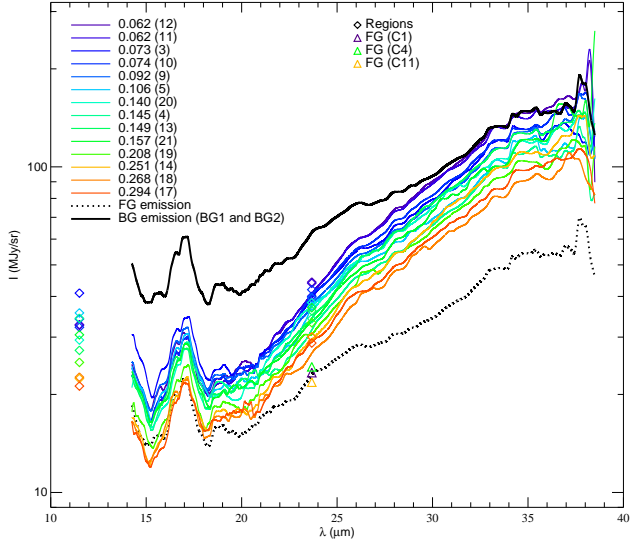}\\
\end{array}$
\end{center}
\caption{\footnotesize 
Extracted spectra from the 10$\arcsec\times$10$\arcsec$ dark regions
(colored lines) and the foreground/background model (dotted/solid
black lines). The comparison between the foreground spectrum and
spectra of $\Sigma_{\rm 24\mu m}\ga$0.25~g~cm$^{-2}$ regions indicates
we see saturation at $\lambda\la$17$\micron$.  Triangles indicate
24$\micron$ photometric data of the saturated cores C1, C4, C11 that
help to normalize the foreground model (also shown in
\ref{fig:bgfgcomp1}). Diamonds are {\it Spitzer}-MIPS 24$\micron$
photometric data points for all regions, which helped normalize the
zodiacal subtraction.}
\label{fig:deredden1}
\end{figure}

The uncertainties of $I_{\nu,0}$ are constrained based on the observed
color variations over the $12\:\micron$ and $24\:\micron$ maps, the
observed MIPS $24\:\micron$ intensity fluctuations in the surrounding
regions outside the IRDC C ellipse, and the intrinsic
photometric errors (that were shown in Fig.~\ref{fig:flucori}).
By considering the scatter in pixel by pixel values of $I_{\rm 12\mu
  m}/I_{\rm 24\mu m}$ in an elliptical annulus 0.7 to 1.3$\times$ the semimajor
axis of the IRDC, we derive a 1$\sigma$ uncertainty, i.e., dispersion,
of about 7\% in the spectral slope of our estimate of the background
spectrum. The color variation of the wavelength range between 24 and
$38\:\micron$ is uncertain and so is potentially larger. Thus, to be
conservative, we assume an uncertainty in the spectral slope
in this wavelength range of $I_{\nu,0}$ of 20\%.
The fluctuations in $I_{\rm 24\mu m}$ are also estimated in the same
annulus and found to have a 1$\sigma$ dispersion of about 18\%, which
we adopt for our error analysis.
Thus the overall uncertainties in $I_{\nu,0}$, including photometric
errors, are summed and are shown in the panels of
Fig.~\ref{fig:bgfluc1} as the grey shaded regions. These panels also
show the scaled foreground and the zodi-corrected observed spectra for
each $24\:\micron$-dark region. These spectra are also shown together
in Figure~\ref{fig:deredden1}. These are the required quantities to
estimate the MIR to FIR extinction law.

\section{Results}\label{S:results}

\subsection{Spectrum of the Galactic plane diffuse ISM}

Figure~\ref{fig:bgfgcomp1} shows that our estimates of the spectrum of
the SED of the diffuse ISM generally agree well will the DL07
model. Our derived spectral slope of this SED from the {\it
  Spitzer}-IRS data rises somewhat more steeply towards longer
wavelengths than the DL07 model. This could either indicate a real
difference or may be due to systematic errors introduced in our
analysis, such as the de-reddening method, which assumed the OH94 thin
ice mantle model opacities.

As an independent check on our derived spectrum of the diffuse ISM, we
can search for evidence of saturation at the short wavelength end of
the observed spectra in high $\Sigma$
regions. Figure~\ref{fig:bgfluc1} and \ref{fig:deredden1} shows
evidence for such saturation in the four highest $\Sigma$ regions (all
with $\Sigma_{\rm 8\mu m}>0.2\:{\rm g\:cm^{-2}}$) at $\lambda\lesssim
18\:{\rm \mu m}$. Note the normalization of the foreground model
(dotted lines in this figure) is based on the $24\micron$ photometry
towards the three saturated cores C1, C4 and C11, so if our model of
emission from the diffuse ISM were in error, then we would not expect
to see saturation, i.e., coincidence of the foreground model with the
observed spectra.

\subsection{The MIR to FIR Extinction Law and Evidence for Grain Growth}   

\begin{figure}
\begin{center}$
\begin{array}{c}
\hspace{0in} \includegraphics[width=3.4in]{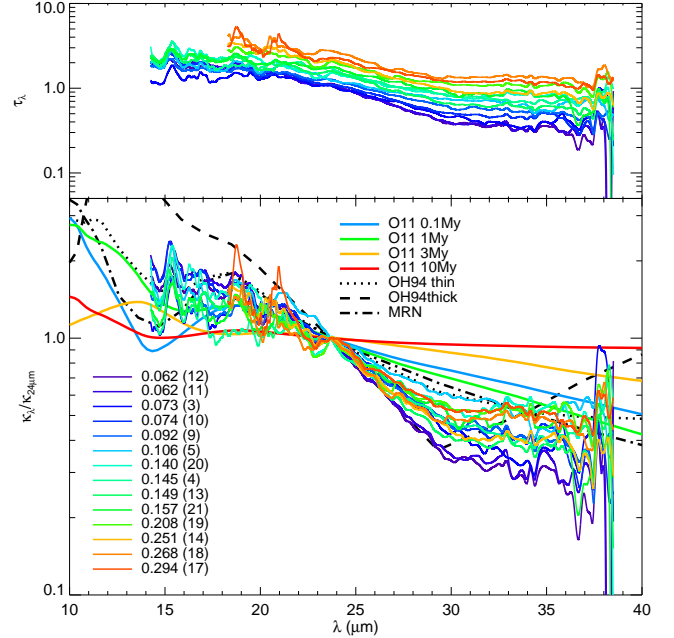}\\
\end{array}$
\end{center}
\caption{\footnotesize 
{\it Upper panel:} The optical depth ($\tau_{\lambda}$) curves of the
14 selected dark positions of the {\it Spitzer}-IRS slit on the IRDC
G028.37+00.07. The color of spectra are same as in
Fig. \ref{fig:deredden1} and are listed in the legend of the lower
panel. {\it Lower panel:} Relative $\kappa$ curves (normalized to
$\kappa_{\rm 24\mu m}$) of the same regions are compared to model
relative opacities of OH94 thin (black dotted line) and thick ice mantle
models (black dashed line), 
OH94 bare grain model (dot-dashed line) 
and O11 models of 0.1, 1, 3 and 10~Myr of coagulation time (blue,
green, gold and red solid lines).
}
\label{fig:relk1}
\end{figure}

\begin{figure*}
\begin{center}$
\begin{array}{c}
\hspace{-0.1in} \includegraphics[width=5.5in]{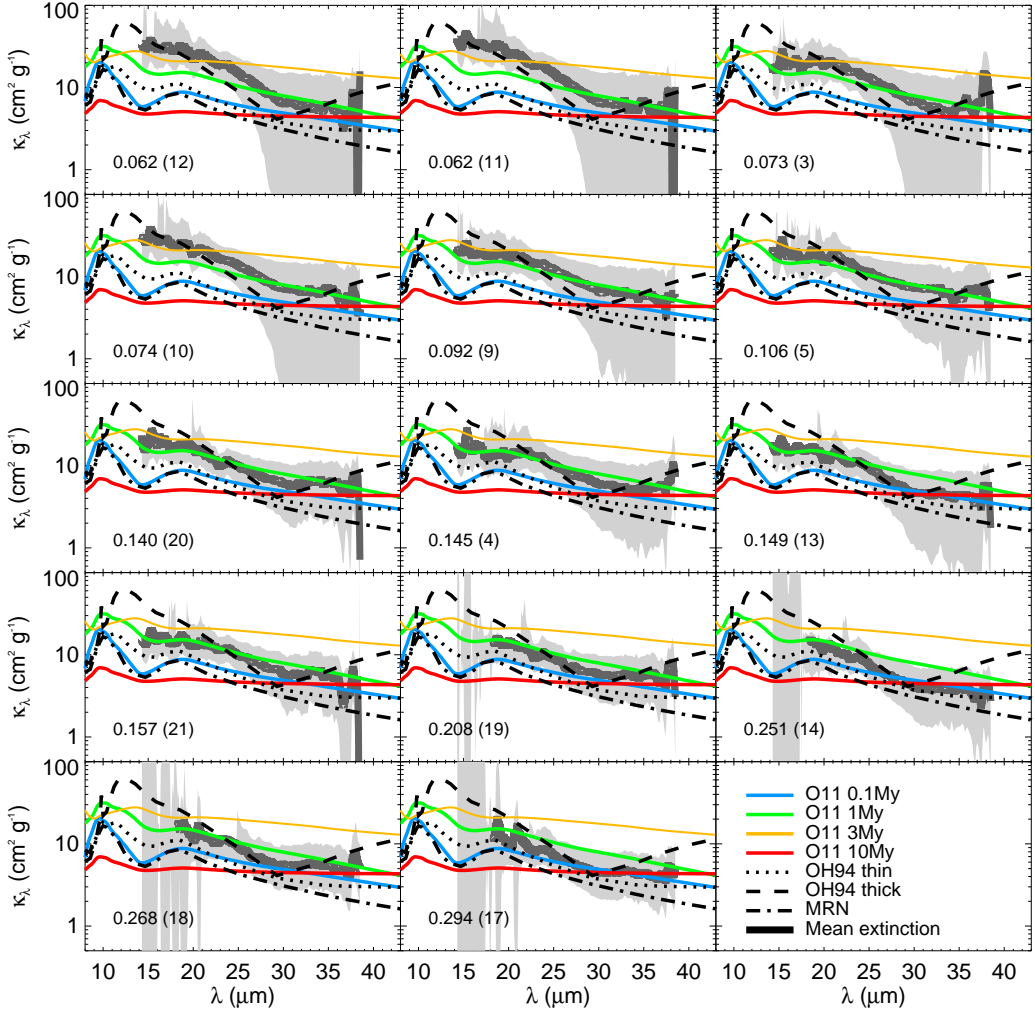}\\
\end{array}$
\end{center}
\caption{\footnotesize 
The opacity ($\kappa_{\lambda}$) curves (thick solid grey line; shaded
areas show uncertainties) of the individual regions (each panel lists
$\Sigma_{\rm 24\mu m}$ in g~cm$^{-2}$ and region number in
parentheses),
compared to model $\kappa_{\lambda}$ of OH94 thin ice mantle (dotted),
thick ice mantle (dashed) and bare grain (MRN) model (dot-dashed).
The O11 models of 0.1, 1, 3 and 10~Myr of coagulation time (blue,
green, gold and red solid lines) are also shown. 
}
\label{fig:absk1}
\end{figure*}

We calculate the optical depth, $\tau_\nu$, of each selected
10$\arcsec\times$10$\arcsec$ 24~$\rm \mu m$-dark region via
$\tau_{\nu}$=ln($I_{\nu,0}$/$I_{\nu,1}$), and these are shown in the
upper panel of Fig.~\ref{fig:relk1}. As expected, higher $\Sigma_{\rm
  24\mu m}$ regions show higher optical depths.
Using $\Sigma_{\rm 24\mu m}$ (from the LT14 FIREX map), we then derive
$\kappa_\nu = \Sigma_{\rm 24\mu m}/\tau_\nu$, and show
$\kappa_\lambda/\kappa_{\rm 24\mu m}$ in the lower panel of
Fig.~\ref{fig:relk1}, along with the OH94 and O11 dust models. The
absolute values of $\kappa_\lambda$ are shown in Fig.~\ref{fig:absk1}.

\begin{figure}
\begin{center}$
\begin{array}{c}
\hspace{0in} \includegraphics[width=3.4in]{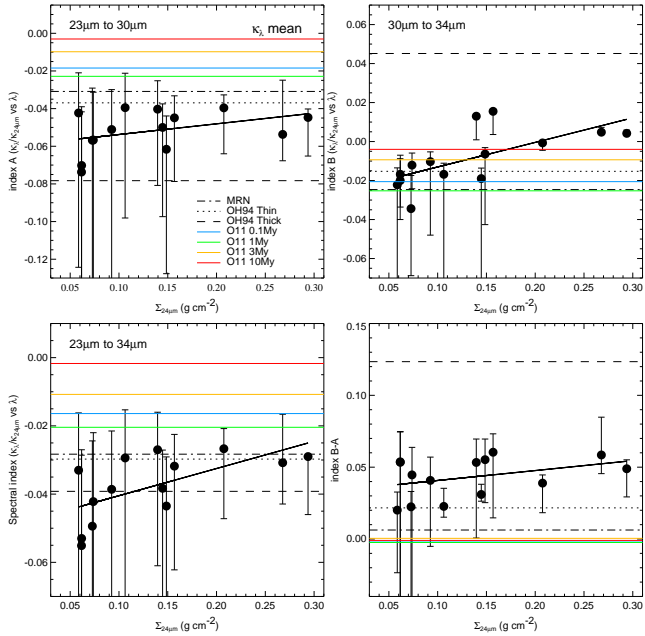}\\
\end{array}$
\end{center}
\caption{\footnotesize 
Spectral indices of extinction laws versus $\Sigma_{\rm 24\mu
  m}$. {\it (a) Top left:} ``Index A,'' the spectral slope, $\beta_A$
(where ${\kappa_\lambda}/{\kappa_{\rm 24\mu m}} \propto
\lambda^{\beta_A}$), measured from 23 to 30~$\rm \mu m$. The black
circles show the results of the fit to the 14 analyzed 24~$\rm \mu
m$-dark regions, with best-fit relation shown by thick solid
line. $\beta_A$ values of the OH94 and O11 dust models are marked with
horizontal lines.
{\it (b) Top right:} Same as (a), but now for ``Index B,'' which is
evaluated from 30 to 34~$\rm \mu m$. Note the relatively large value
of the OH94 thick ice mantle model, due to the water ice feature.
{\it (c) Bottom left:} Same as (a), but now for ``Index C,'' which is
evaluated from 23 to 34~$\rm \mu m$.
{\it (d) Bottom right:} $\beta_B - \beta_A$ versus $\Sigma_{\rm 24\mu
  m}$.
}
\label{fig:index1}
\end{figure}

To quantify the shape of the extinction laws that are probed by the
{\it Spitzer}-IRS data we fit various spectral indices, $\beta$, where
${\kappa_\lambda}/{\kappa_{\rm 24\mu m}} \propto \lambda^{\beta}$). We
define Index A as $\beta_A$ being measured from 23 to 30~$\rm \mu m$;
Index B as $\beta_B$ being measured from 30 to 34~$\rm \mu m$; and
Index C as $\beta_C$ being measured from 23 to 34~$\rm \mu m$. The
divisions of A and B at 30~$\rm \mu m$ were chosen to accentuate the
``V''-shaped water ice feature that is prominent in the thick ice
mantle models of OH94. Figure~\ref{fig:index1} shows these various
indices and the combination of $\beta_B-\beta_A$ as a function of
$\Sigma_{\rm 24\mu m}$. The uncertainties of the spectral indices of
each region are approximately estimated by adopting the range of
indices that are allowed from the upper and lower limits of the
SIREX-derived opacities (i.e., from the shaded areas of
Figure~\ref{fig:absk1}).

The absolute values of these indices and the combination
$\beta_B-\beta_A$ are consistent with the OH94 thin ice mantle model.
The values of $\beta_B$ disfavor the thick ice mantle model. O11
models with $\sim 1$~Myr of coagulation do reasonably well, but in
general the O11 models are too flat. However, we caution that there
are systematic errors (e.g., the dereddening of the background
regions) that could be affecting the derived spectral indices of the
extinction laws.

\begin{figure*}
\begin{center}$
\begin{array}{c}
\hspace{0in} \includegraphics[width=5.5in]{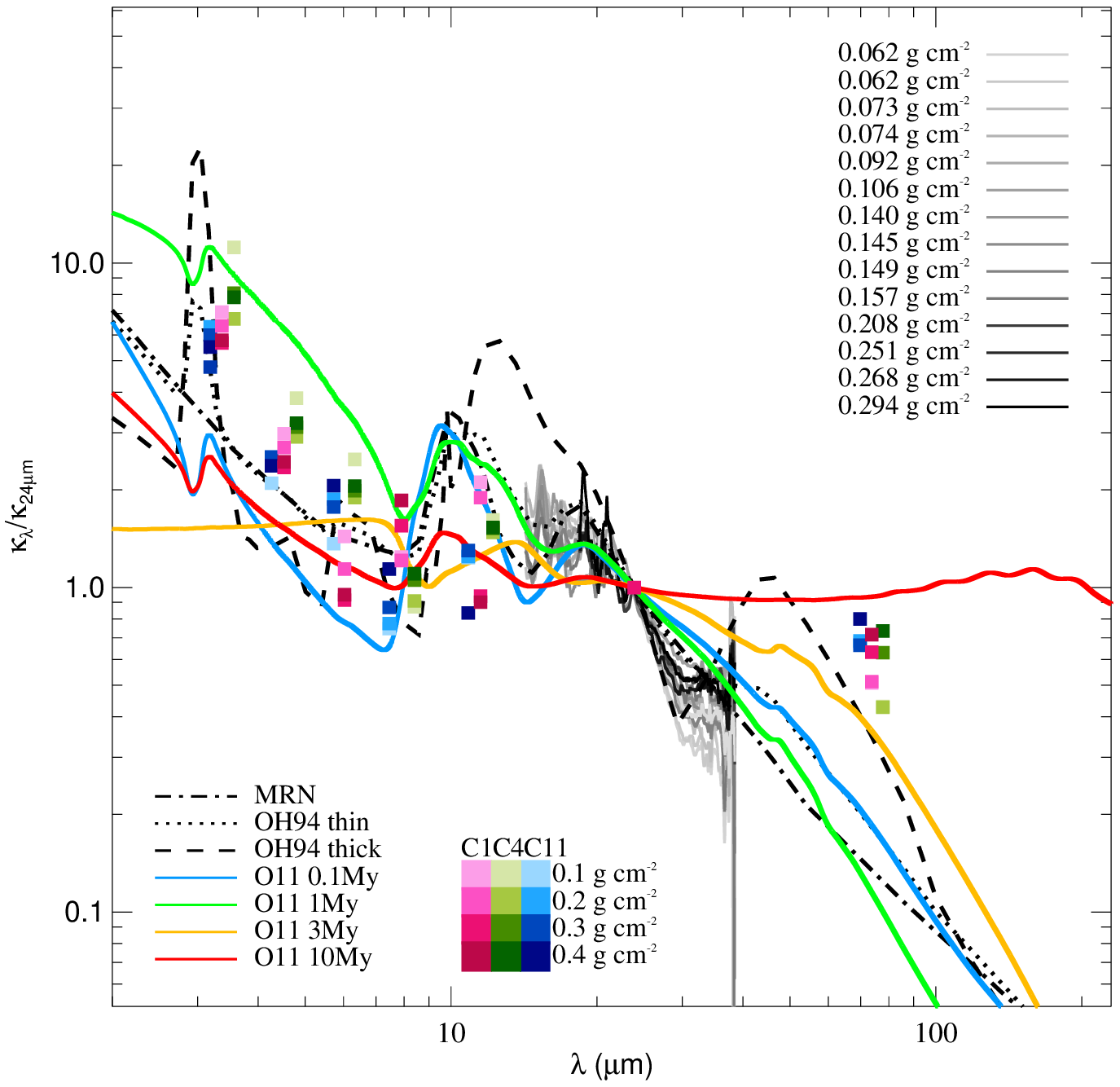}\\
\end{array}$
\end{center}
\caption{\footnotesize
Overall 24~$\rm \mu m$-normalized MIR to FIR extinction law, combining
photometric (MIREX and FIREX) results of \S\ref{S:photometry} (from
Fig.~\ref{fig:photometry}d) and spectroscopic (SIREX) results.
The solid lines with black-white gradient indicate SIREX
$\kappa_{\lambda}/\kappa_{\rm 24\mu m}$ results at different $\Sigma_{\rm 24\mu m}$.
}
\label{fig:summary}
\end{figure*}

Next we look for systematic variation of the extinction laws with mass
surface density. All three indices and the combination
$\beta_B-\beta_A$ show hints of flattening as $\Sigma_{\rm 24\mu m}$
increases. The Pearson $p$ values are reasonably significant for Index
B and C ($\sim$0.008 and $\sim$0.01, respectively).
Thus Index B shows relatively significant changes with increasing
$\Sigma_{\rm 24\mu m}$, which may indicate that water ice is in fact
beginning to change the shape of the extinction curves in the denser
regions (although not to the extent of OH94 thick ice mantle
models). Such an interpretation would help to reconcile the results of
the {\it Spitzer}-IRS extinction curves with the photometric results
of \S\ref{S:photometry}, which we present together in a summary in
Figure~\ref{fig:summary}.

\section{Conclusions}

We have presented a new method of MIR to FIR spectroscopic extinction
(SIREX) mapping of an IRDC, which can observationally trace grain
growth processes in a region that is expected to form massive stars
and a massive star cluster. This is the first study to determine the
detailed extinction law in the wavelength range $14\:\micron$ to
$38\:\micron$ in such dense regions. In combination with our MIREX and
FIREX maps, we are also able to search for systematic variation in the
extinction law as a function of mass surface density.

Our new results, especially from {\it WISE} 12~$\rm \mu m$ MIREX
mapping and {\it Spitzer}-IRS SIREX mapping, indicate that
$\kappa_\lambda/\kappa_{\rm 24\mu m}$ is relatively flat from $\sim 5$
to $\sim 25\:{\rm \mu m}$. The strong water ice absorption feature of
the OH94 thick ice mantle models is not seen. SIREX results then
indicate a relatively steep drop of $\kappa_\lambda/\kappa_{\rm 24\mu
  m}$ out to $\sim 30\:{\rm \mu m}$, but then hints of some
flattening, especially in denser regions, out to $\sim 35\:{\rm \mu
  m}$. The FIREX results at $70\:{\rm \mu m}$ suggest this flattening
extends out to these long wavelengths and is also stronger in higher
density environments. Comparison with OH94 and O11 dust models
suggests that this may require some combination of water ice mantle
growth and grain coagulation.

One implication of these results is that there could be very
significant grain growth on large scales within IRDCs, so that
pre-stellar cores should be modeled with such large grains. This can
potentially affect cooling rates (generally leading to cooler
equilibrium temperatures; e.g., Keto \& Caselli 2008).
It can also affect astrochemistry and ionization fractions.

Another important implication is for mass estimates of dense regions
from sub-mm and mm dust emission. The difference in derived mass
surface density that results from using either thin or thick OH94 dust
opacities is about a factor of three.

Future studies of spectroscopic observation toward dense cores and
clumps by using finer angular resolution and higher sensitivity (e.g.,
{\it JWST}-MIRI) will help to better explore the evidence of grain
growth in different star-forming environments.

\acknowledgements
We thank to E. Schisano for WISE, Herschel data and discussions, and M. J. Butler, B. T. Draine, J. G. Ingalls, S. Kong,  A. Li and R. Paladini for discussions. The comments of an anonymous referee helped to improve the article. JCT acknowledges support from NASA Astrophysics Data Analysis Program grants ADAP10-0110 and ADAP14-0135.

\end{document}